# Early-detection and classification of live bacteria using time-lapse coherent imaging and deep learning

Hongda Wang[1,2,3,†], Hatice Ceylan Koydemir[1,2,3,†], Yunzhe Qiu[1,2,3,†], Bijie Bai[1,2,3], Yibo Zhang[1,2,3], Yiyin Jin[1], Sabiha Tok[1,2,3,4], Enis Cagatay Yilmaz[1], Esin Gumustekin[5], Yair Rivenson[1,2,3], Aydogan Ozcan[1,2,3,6,*]

[1]Electrical and Computer Engineering Department, University of California, Los Angeles, CA, 90095, USA

[2]Bioengineering Department, University of California, Los Angeles, CA, 90095, USA

[3]California NanoSystems Institute, University of California, Los Angeles, CA, 90095, USA

[4]Department of Biophysics, Istanbul Medical Faculty, Istanbul University, Istanbul, 22000, Turkey

[5]Department of Microbiology, Immunology, and Molecular Genetics, University of California, Los Angeles, CA, 90095, USA

[6]Department of Surgery, David Geffen School of Medicine, University of California, Los Angeles, CA, 90095, USA.

†Equal contributing authors.

*Corresponding author: ozcan@ucla.edu

**Abstract:**

Early identification of pathogenic bacteria in food, water, and bodily fluids is very important and yet challenging, owing to sample complexities and large sample volumes that need to be rapidly screened. Existing screening methods based on plate-counting or molecular analysis present some tradeoffs with regards to the detection time, accuracy/sensitivity, cost, and sample preparation complexity. Here we present a computational live bacteria detection system that periodically captures coherent microscopy images of bacterial growth inside a 60-mm-diameter agar-plate and analyzes these time-lapsed holograms using deep neural networks for rapid detection of bacterial growth and classification of the corresponding species. The performance of our system was demonstrated by rapid detection of *Escherichia coli* and total coliform bacteria (i.e., *Klebsiella aerogenes* and *Klebsiella pneumoniae* subsp. *pneumoniae*) in water samples. These results were confirmed against gold-standard culture-based results, shortening the detection time of bacterial growth by >12 h as compared to the Environmental Protection Agency (EPA)-approved analytical methods. Our experiments further confirmed that this method successfully detects 90% of bacterial colonies within 7–10 h (and >95% within 12 h) with a precision of 99.2-100%, and correctly identifies their species in 7.6–12 h with 80% accuracy. Using pre-incubation of samples in growth media, our system achieved a limit of detection (LOD) of ~1 colony forming unit (CFU)/L in ≤ 9 h of total test time. This computational bacteria detection and classification platform is highly cost-effective (~$0.6 per test) and high-throughput with a scanning speed of 24 cm$^2$/min over the entire plate surface, making it highly suitable for integration with the existing analytical methods currently used for bacteria detection on agar plates. Powered by deep learning, this automated and cost-effective live bacteria detection platform can be transformative for a wide range of applications in microbiology by significantly reducing the detection time, also automating the identification of colonies, without labeling or the need for an expert.

**Introduction**

Rapid and accurate identification of live microorganisms is of great importance for a wide range of applications[1–8], including drug discovery screening assays[1–3], clinical diagnoses[4], microbiome studies[5,6], and food and water safety[7,8]. Waterborne diseases affect more than 2 billion people worldwide[9], causing substantial economic burden; for example, treatment of waterborne diseases costs more than $2 billion annually in the United States (US) alone, with 90 million cases recorded per year[10].

Among waterborne pathogen-related problems, one of the most common public health concerns is the presence of total coliform bacteria and *Escherichia coli* (*E. coli*) in drinking water, which indicates fecal contamination. Analytical methods used to detect *E. coli* and total coliforms are based on culturing the obtained samples on solid agar plates (e.g., the US Environmental Protection Agency (EPA) 1103.1 and EPA 1604 methods) or in liquid media (e.g., Colilert test), followed by visual recognition and counting by an expert, as described in the EPA guidelines[11–13]. While the use of liquid growth media for the detection of fecal coliform bacteria provides high sensitivity and specificity, it requires at least 18 h for the final read-out. The use of solid agar plates is a relatively more cost-effective method and provides flexibility for the volume of the sample to be analyzed, which can vary from 100 mL to several liters by using a membrane filtration technique to enhance sensitivity. However, this traditional culture-based detection method requires the colonies to grow to a certain macroscopic size for visibility, which often takes 24–48 h in the case of bacterial samples. Alternatively, molecular detection methods[14,15] based on e.g., the amplification of nucleic acids can reduce the assay time to a few hours, but they generally lack the sensitivity for detecting bacteria at very low concentrations, e.g., 1 colony-forming unit (CFU) per 100-1000 mL, and are *not* capable of differentiating between live and dead microorganisms.[16] Furthermore, there is *no* EPA-approved nucleic acid-based analytical method[17] for detecting coliforms in water samples.

All in all, there is a strong and urgent need for an automated method that can achieve rapid and high-throughput colony detection with high sensitivity (routinely achieving e.g., 1 CFU per 100-1000 mL in less than 12 h), to provide a powerful alternative to the currently available EPA-approved gold-standard analytical methods that (1) are slow, taking ~24–48 h, and (2) require experts to read and quantify samples. To address this important need, various other approaches have been investigated for the detection of total coliform bacteria and *E. coli* in water samples, including solid phase cytometry[18], droplet based micro-optical lens arrays[19], fluorimetry[20], luminometry[21], and fluorescence microscopy[22]. Despite the fact that these methods provide high sensitivity and some time savings, they cannot handle large sample sizes (e.g., ≥100 mL) or cannot perform automated classification of bacterial colonies.

To provide a highly-sensitive and high-throughput system for early detection and classification of live microorganisms and colony growth, here we present a time-lapse coherent imaging platform that uses two different deep neural networks (DNNs) for its operation. The first DNN is used to detect bacterial growth as early as possible, and the second one is used to classify the type of growing bacteria, based on the spatio-temporal features obtained from the coherent images of an incubated agar-plate (see Fig. 1). In this live bacteria detection system, which is integrated with an incubator, lens-free holographic images of the agar-plate sample are captured by a monochromatic complementary metal–oxide–semiconductor (CMOS) image sensor that is mounted on a translational stage. The system rapidly scans the entire area of two separate agar plates (~ 56.52 cm$^2$) every 30 min, and utilizes these time-resolved holographic images for accurate detection, classification, and counting of the growing colonies as early as possible (see Fig. 2a). This unique system enables high-throughput periodic monitoring of an incubated sample by scanning a 60-mm diameter agar-plate in 87 s with an image resolution of <4 μm; it continuously calculates differential images of the sample of interest for the early and accurate detection of bacterial growth. The spatio-temporal features of each non-static object on the plate are continuously analyzed using deep learning to yield the count of bacterial growth, and to automatically identify the type(s) of bacteria growing on different parts of the agar-plate.

We demonstrated the efficacy of this platform by performing early detection and classification of three types of bacteria, i.e., *E. coli*, *Klebsiella aerogenes (K. aerogenes)*, and *Klebsiella pneumoniae (K. pneumoniae)*, and achieved a limit of detection (LOD) of ~1 CFU/L in ≤ 9 h of total test time. Moreover, we achieved detection time savings of more than 12 h as compared to the gold-standard EPA methods[23], which usually require at least 24 h to obtain a result. We also quantified the growth statistics of these three different species and provided a detailed growth analysis of each type of bacteria over time. Our detection and classification neural network models were built, trained and validated with ~16,000 individual colonies resulting from 71 independent experiments and were blindly tested with 965 individual colonies collected from 15 independent experiments that were never used in the training phase. In our blind testing, the trained models demonstrated 80% detection sensitivity within 6–9 h, 90% detection sensitivity within 7–10 h, and >95% detection sensitivity within 12 h, while maintaining ~99.2-100% precision at any time point after 7 h, also achieving correct identification of 80% of all three the species within 7.6–12 h. In terms of species-specific accuracy of our classification network, within 12 h of incubation we achieved ~97.2%, ~84.0%, and ~98.5% classification accuracy for *E. coli*, *K. aerogenes*, and *K. pneumoniae*, respectively. These results confirm the transformative potential of our platform, which not only enables highly sensitive, rapid and cost-effective detection of live bacteria (with a cost of $0.6 per test), but also provides a powerful and versatile tool for microbiology research.

**Results**

We demonstrated our system by monitoring bacterial colony growth within 60-mm-diameter agar-plates, and quantitatively analyzed the capabilities of the platform for early detection of bacterial growth and classification of bacterial species. To demonstrate its proof-of-concept, we aimed to automatically detect, classify, and count *E. coli* and coliform bacteria in water samples using our deep learning-based platform. Throughout our training and blind testing experiments, we used water suspensions spiked with coliform bacteria, including *E. coli*, *K. aerogenes*, and *K. pneumoniae*, and chlorine stressed *E. coli*. A chromogenic agar medium designed for the specific detection and counting of *E. coli* and other coliform bacteria in food and water samples was used as a culture medium for specificity (see the Methods section for details). This chromogenic medium results in blue color for *E. coli* colonies and mauve color for the colonies of other coliform bacteria (e.g., *K. aerogenes* and *K. pneumoniae)*. Additionally, it inhibits the growth of different bacteria (e.g., *Bacillus subtilis*), or yields colorless colonies in the presence of other bacteria in the sample[24].

Following the sample preparation method illustrated in Fig. 2a, the sample is placed inside the lens-free imaging system with the agar surface facing the image sensor. After an initialization step, the platform automatically captures time-lapsed holographic images of two separate Petri dishes (covering a total sample area of 28.26 x 2 = 56.52 cm$^2$) every 30 min over a duration of 24 h starting from the incubation time; these individual holograms are digitally stitched together and rapidly reconstructed to reveal the bacterial growth patterns on the agar surface (see the Methods section). The reconstructed images of the sample captured at different time points are computationally processed using a differential image analysis method to automatically detect and classify bacterial growth and colonies using two different trained DNNs (see Fig. 3), which will be detailed next.

**Design and training of neural networks for bacterial growth detection and classification**

We designed a two-step framework for bacterial growth detection and classification. The first step selects colony candidates with differential image analysis and refines the results with a detection DNN. We designed a pseudo-3D (P3D) DenseNet[25] architecture to process our complex-valued (i.e., phase and amplitude) time-lapse image stacks (see the Methods section). In each time-lapse imaging experiment, we used 4 time-consecutive frames (4 × 0.5 = 2 h) as a running window for differential image analysis to extract individual regions-of-interest (ROIs) containing objects that changed their amplitude and/or phase signatures as a function of time. These initially-detected objects that were extracted by the differential analysis algorithm were either growing colonies or surface impurities, e.g., from spreading the sample on the agar surface, evaporation of air bubbles in the agar plate, or coherent light speckles. We then used a DNN-based detection model to eliminate non-bacterial objects, and only kept the growing colonies (i.e., the true positives), as illustrated in Fig. 2b. We used sensitivity (or true positive rate, TPR) and precision (or positive predictive value, PPV) measurements to quantify our results. Sensitivity is defined as:

$$TPR = TP / P,$$

where TP refers to the number of true positive predictions from our system, and P refers to the total number of colonies resulting from manual plate-counting *after* 24 h (i.e., the ground truth). Precision is defined as:

$$PPV = TP / (TP + FP),$$

where FP refers to the number of false positive predictions from our system.

In total 13,712 growing colonies (*E. coli*, *K. aerogenes*, and *K. pneumoniae*) and 30,000 non-colony objects captured from 66 separate agar plates were used in the training phase. Another 2,597 colonies and 13,078 non-colony objects from 5 independent plates were used as validation dataset to finalize our network models, and achieved a TPR of ~95% and a PPV of ~95% once the network converged, which took ~ 4 h of training time. Examples of the training loss and detection accuracy curves are shown in Supplementary Fig. S1.

The second step further classifies the species of the detected colonies with a classification DNN model following a similar network architecture. To accommodate different growth rates of bacterial colonies, we use a longer time window in this classification neural network, containing 8 consecutive frames (8 × 0.5 = 4 h) for each sub-ROI. Since the bacterial growth detection network uses a shorter running time window of 2 h, there is a natural 2-hour time delay between the successful detection of a growing colony and the classification of its species. The network was trained with 7,919 growing colonies, which contained 3,362 *E. coli*, 1,880 *K. aerogenes*, and 2,677 *K. pneumoniae* colonies, and it was validated with 340 *E. coli*, 205 *K. aerogenes*, and 988 *K. pneumoniae* colonies from 6 independent plates, and reached a validation classification accuracy of ~89% for *E. coli*, ~95% for *K. aerogenes*, and ~98% for *K. pneumoniae* when the network model converged (Supplementary Fig. S2).

After these network models were finalized through the training and validation data, we tested their generalization capabilities with an additional set of experiments that were never seen by the networks before; the results of these blind tests are detailed next.

**Blind testing results on early detection of bacterial growth**

First, we blindly tested the performance of our system in early detection of bacterial colonies with 965 colonies from 15 plates that were not presented during the network training or validation stages. We compared the predicted number of growing colonies on the sample within the first 14 h of incubation against a ground truth colony count obtained from plate counting *after* 24 h of incubation time. Each of the 3 sensitivity curves (Figs. 4a–4c) were averaged across repeated experiments for the same species, e.g., 4 experiments for *K. pneumoniae*, 7 experiments for *E. coli*, and 4 experiments for *K. aerogenes*, so that each data point was calculated from ~300 colonies. The results demonstrated that our system was able to detect 80% of true positive colonies within ~6.0 h of incubation for *K. pneumoniae*, ~6.8 h of incubation for *E. coli*, and ~8.8 h of incubation for *K. aerogenes*, respectively. It further detected 90% of true positives after ~1 additional hour of incubation, and >95% of the true positive colonies of all of the 3 species within 12 h. The results also reveal that the early detection sensitivities in Figs. 4a–4c are dependent on the length of the lag phase of each tested bacteria species, which demonstrates inter-species variations. For example, *K. pneumoniae* started to grow earlier and faster than *E. coli* and *K. aerogenes*, whereas *K. aerogenes* did not reach to a detectable growth size until 5 h of incubation. Furthermore, when the tails of the sensitivity curves are examined, one can find out that some of the *E. coli* colonies showed late "wake-up" behavior, as highlighted by the purple arrow in Fig. 4b. Although most of the *E. coli* colonies were detected within ~10 h of incubation time, some of them did not emerge until ~11 h after the start of the incubation phase.

We also quantified the false positive rate of our platform with the PPV curve as shown in Fig. 4d, which was averaged across all the experiments covering all the species, i.e., 965 colonies from 15 agar-plates. The precision can be low at the beginning of the experiments (the first 4 h of incubation), because the number of detected true positive colonies is very small, especially for *K. aerogenes*. This means even a single false positive-detected colony can dramatically affect the precision calculation. Nevertheless, the precision quickly rises up to ~100% within 6 h of incubation and is maintained at 99.2-100% for all the tested species after 7 h of incubation.

We should emphasize here that the results presented in Fig. 4 represent the *lower limits* of the detection capabilities of our system since we calculated these sensitivities with regard to the number of true positive colonies *after* 24 h of incubation, whereas some of these colonies actually did *not* exist at the early stages due to delayed growth; stated differently in some cases, there were no colonies present at the early stages of the incubation period. We also note that the rising sensitivity curves in our results stand for the emergence of new bacterial colonies, in addition to the growth of colonies. Even though the sensitivity curves converge to flat lines after 12 h, the colonies keep growing exponentially until much later. Therefore, our system detects emerging colonies at an early stage, when they first appear, forming micro-scale features invisible to the naked eye.

These observations also indicate that our system can be very effective and used for high-throughput quantitative studies to better understand microorganism behavior under different conditions, such as the evaluation of the differences in growth rates between stressed bacteria (e.g., under nutrient deprivation or chlorine treatment) and normal bacteria.[26–30] There are several reasons to detect and enumerate chlorine stressed or injured coliform bacteria. First of all, the detection of injured *E.coli* or total coliform bacteria is directly related to the sensitivity of the detection platform.[30] For an effective and sensitive detection platform, the false negative results should be avoided for public health safety. Another important reason is that the detection of injured *E.coli* or low numbers of *E.coli* in water samples is correlated to Salmonella outbreaks, a foodborne pathogen causing 1.2 million illnesses and ~500 deaths per year in the US[31], which forms an indirect indicator of contamination in irrigation water.[32] To evaluate the capabilities of our system to detect injured bacteria, we prepared and imaged 3 agar plates containing chlorine-stressed *E.coli* (see the Methods section), and characterized their growth using our detection workflow as summarized in Fig. 4e. Our results indicate that we can detect colony formation for chlorine-stressed *E. coli* on average with a ~2 h delay compared to the regular *E. coli* strain.

**Blind testing results on classification of growing bacteria**

In addition to providing significant detection time savings while also achieving a very good sensitivity and precision for early detection of bacterial growth, our method also provides automated classification of the corresponding species of the detected bacteria using a trained neural network. Therefore, an additional advantage of our system is its capability to further classify total coliform sub-species, which is not possible with traditional agar-plate counting methods. For example, both *K. pneumoniae* and *K. aerogenes* colonies appear mauve in our agar-plates. However, since our classification neural network does not only rely on the byproducts of the colorimetric reactions, it can successfully distinguish between different species based on their unique spatiotemporal growth signatures acquired by our platform at the micro-scale.

Fig. 5 shows our blind testing results on species classification using the same experiments reported in the blinded early-detection tests, containing 965 colonies of 3 different species from 15 agar-plates. In these results, if a colony had not been detected in the previous step (i.e., a false negative event compared to the 24 h reading), it was naturally not sent to the classification neural network. We defined the recovery rate as the number of colonies correctly classified into their corresponding species using our system, divided by the total number of colonies counted *after* 24 h. As the classification of each individual colony is an independent event, we calculated the recovery rate for each bacteria species (reported in Figs. 5a–5c) using all of the colonies detected in the previous step, i.e., 336, 280, and 339 colonies of *E. coli*, *K. aerogenes*, and *K. pneumoniae,* respectively. The shaded area in each curve represents the *highest* and *lowest* recovery rates found in all the corresponding experiments at each time point. The classification neural network correctly classified ~80% of all of the colonies within ~7.6 h, ~8 h, and ~12 h for *K. pneumoniae*, *E. coli*, and *K. aerogenes*, respectively. We once again emphasize that the results presented in Fig. 5(a-c) represent the lower limits of the classification capabilities of our system since the ground truth is acquired after 24 h of incubation. In reality, at various earlier time points within the incubation period, there was no growth for certain regions of the plates, which exhibited significantly delayed growth. To further demonstrate the classification performance of our trained neural network in a manner that is decoupled from the sensitivity of the previous detection network, we report the classification confusion matrix in Fig. 5d for all the colonies that were sent to the classification network for blind testing at 12 h after the start of the incubation. The trained network achieved classification accuracies of ~97.2%, ~84.0%, and ~98.5% for *E. coli*, *K. aerogenes*, and *K. pneumoniae*, respectively.

**Limit of detection as a function of the total test time**

We further quantified the detection limit of our system and compared its performance against both Colilert® 18, which is an EPA approved method, and traditional plate counting (Supplementary Table S1 and Supplementary Figure S3). To make up for the CFU loss during the sample transfer from the water suspension to the filter membrane, we introduced a signal amplification step by pre-incubating the water sample under test, mixing it with a growth medium for 5 h at 35 °C before the filtration step (see the Methods section for details). For each measurement, 2 agar plates were prepared and monitored at the same time for comparison, one of which was for the sample amplified with 5-h pre-incubation step before the filtering, while the other one was for the sample directly filtered and transferred to the agar plate (see Supplementary Figure S3). Both plates were incubated for the same amount of time at each imaging time point to provide a fair comparison between the two. The measurements were repeated using different concentrations of *E. coli* suspensions; these concentrations were compared to the average of three replicates of the same samples prepared using the Colilert®-18 method (Supplementary Figure S3). As shown in Fig. 6a, our system is able to surpass the sensitivity of Colilert®-18 within ~8 h in total (including the time for signal amplification, sample concentration, and time-lapse imaging, altogether) and reach >2 times the sensitivity of Colilert®-18 in ~9 h. We also quantified the LOD of our system by preparing and imaging 3 agar plates without bacteria, which show on average <1 CFU count from our setup throughout the test period from 5 h to 14.5 h (Fig. 6c), revealing a detection limit of $\mu + 3\sigma =$ ~2 CFU per test, where $\mu$ and $\sigma$ refer to the mean and standard deviation of the detected CFU count, respectively. Due to the effective signal amplification enabled by the pre-incubation step, even with the lowest bacterial concentration of ~1 CFU/L, our system was able to detect 2 CFU at 8.5 h, and 12 CFU at 9 h; in comparison, for the same contaminated water sample Colilert® 18 achieved 1.4±1.6 CFU/L after 18 hours of incubation. Furthermore, for all the concentrations we have experimented with (~1-160 CFU/L), our system successfully detected more than 2 CFU per test in ≤ 9 h of test time, including all the necessary steps i.e., the time for signal amplification, sample concentration, and time-lapse imaging; these results reveal that our system with a pre-incubation step achieves a detection limit of ~1 CFU/L within ≤ 9 h of total test time.

We also observe in Fig. 6b that without the signal amplification enabled by pre-incubation, the detection performance is negatively affected due to the low transfer rate of bacteria from the container to the agar plate (also see Supplementary Figure S4). In general, the sensitivity and LOD of our method might be further improved by increasing the pre-incubation time of the water-broth mixture, at the cost of an increase in the total time to achieve automated detection and classification.

**Discussion**

We demonstrated a new platform for early detection and classification of bacterial colonies, which is fully compatible with the existing EPA-approved methods and can be integrated with them to considerably improve the analysis of agar plates[33]. The presented approach can automatically detect bacterial growth as early as in 3 h and can detect 90% of bacterial colonies within 7–10 h (and >95% within 12 h), with a precision of 99.2-100%. The system also correctly classifies ~80% of all of the tested bacterial colonies within 7.6, 8.8, and 12 h, for *K. pneumoniae*, *E. coli*, and *K. aerogenes*, respectively. These results present a total time saving of more than 12 h as compared to the gold-standard methods (e.g., Colilert test and Standard Method 9222 B), which require 18-24 h. In addition to automated detection of live bacteria and species classification, the rich spatio-temporal information embedded in the holographic images can be used for more advanced analysis of water samples and microbiology research in general.

Another advantage of this system is its high-throughput imaging capability of agar plates. Our prototype performs a 242-tile-scan within 87 s per agar plate, corresponding to a raw image scanning throughput of ~49 $cm^2$/min. To leave sufficient data redundancy for image post-processing, we set a relatively large overlap of 30% on each side of the acquired holographic image, which reduces the effective imaging throughput of our platform to ~24 $cm^2$/min. As our system is based on lens-free holographic microscopy, it does not require mechanical axial focusing at each position, and instead auto-focuses onto the object plane computationally. We characterized the spatial resolution of our system by imaging a resolution test target, as shown in Supplementary Fig. S5, achieving a linewidth resolution of ~3.5 µm, roughly equivalent to the performance a 4× objective lens with a numerical aperture (NA) of ~0.1. Compared to our system, which takes 87 s to scan an agar plate, a traditional lens-based bright-field microscope using a 4× objective lens would approximately take ~128 min to scan a plate with the same diameter (60 mm), owing to the requirement for mechanical axial focusing (see Supplementary Table S2).

Another important advantage of our system is the minimum requirement for optical alignment; the presented platform is tolerant towards structural changes, such as variations in the sample-to-sensor distance or the illumination angle. Our computational refocusing capability also enables screening of thick samples, e.g., melted agar-plates.[34] An example of a 3D sample is illustrated in Supplementary Fig. S6, where *E. coli* colonies are formed at different depths inside the solid culture medium having a thickness of ~5 mm. For example, the colony marked with "A" grew at ~2170 µm measured from the surface of the agar, whereas the colony marked with "B" was on the agar surface. Our system localizes colonies growing at different depths within a 3D culture medium using a single hologram measurement at each scanning position. However, it is a non-trivial task to image a 3D sample using a conventional lens-based microscope, because of the time required for mechanical focusing, and the refractive index mismatch between the culture medium and the air, which degrades the image resolution as a result of aberrations. Therefore, the corresponding brightfield microscopy images of the whole plates could only be acquired after 24 h of incubation.

Our platform also employs a modular design which is scalable to a larger sample size and a smaller tile-scan time interval. The monitoring field of view (FOV) of this platform is fundamentally limited by the image acquisition time and the stage moving speed. With further optimization of the hardware and control algorithms, an imaging throughput of >50 $cm^2$/min can be reached. Alternatively, several image sensors can be installed and connected to a single computer for high-throughput parallel imaging.[35] In our proof-of-concept implementation, our image processing for each time interval takes ~20 min, and fits well into our 30 min measurement period between each scan. In case a shorter time interval is desired, an image processing procedure implemented using MATLAB and Python/PyTorch programming environments can be further sped up by programming in C/C++. With the help of graphic processing units (GPUs), one can expect >10-fold time savings in computation.[36]

This unique platform is integrated with an incubator to keep the agar plates at a desired temperature. The incubator is a thermal glass plate which contains uniform lines of optically clear indium tin oxide (ITO) electrode for heating the sample placed on top. It is controlled with a controller, which is lightweight. Throughout the experiments, we set the temperature at the agar surface where bacteria grew at ~38 °C, so that all of the tested bacteria species could grow and develop colonies. This temperature was not optimized to promote the growth of a specific species. Therefore, adjustment of the incubation environment, its temperature and humidity can potentially be used to further accelerate colony growth and help us achieve earlier detection and identification of specific bacterial colonies. Another important parameter for growth of microorganisms is humidity. Our system can also be integrated with a controlled humidity chamber for better control and analysis of growth dynamics of various microorganisms.[37]

In summary, we presented a deep learning-based live bacteria monitoring system for early detection of growing colonies and classification of colony species using deep learning. We demonstrated a proof-of-concept device using 3 types of bacteria, i.e., *E. coli*, *K. aerogenes*, and *K. pneumoniae*, and achieved >12 h time savings for both the early detection and the classification of growing species as compared to the gold standard EPA-approved methods. Achieving an LOD of ~1 CFU/L in ≤9 hours, we believe that this versatile system will not only benefit water and food quality monitoring, but also provide a powerful tool for microbiology research.

## Methods

a. **Sample preparation**

*Safety practices:* We handled all the bacterial cultures and performed all the experiments at our Biosafety Level 2 laboratory, in accordance with the environmental, health, and safety rules of the University of California, Los Angeles.

*Studied organisms:* We used E. coli (Migula) Castellani and Chalmers (ATCC® 25922™) (risk level 1), *K. aerogenes* Tindall *et al.* (ATCC® 49701™) (risk level 1), and *K. pneumoniae* subsp. *pneumoniae* (Schroeter) Trevisan (ATCC®13883™) (risk level 2) as our culture organisms.

*Preparation of poured agar plates*: We used CHROMagar™ ECC (product no. EF322, DRG International, Inc., Springfield, NJ, USA) chromogenic substrate mixture as the solid growth medium for the detection of *E. coli* and total coliform colonies. 8.2 g of CHROMagar™ ECC was mixed with 250 mL of reagent grade water (product no. 23-249-581, Fisher Scientific, Hampton, NH, USA), using a magnetic stirrer bar. The mixture was then heated to 100 °C on a hot plate while being stirred regularly. After cooling the mixture to ~50 °C, 10 mL of the mixture was dispensed into Petri dishes (60 mm × 15 mm) (product no. FB0875713A, Fisher Scientific, Hampton, NH, USA). The agar plates were allowed to solidify, sealed using a parafilm (product no. 13-374-16, Fisher Scientific, Hampton, NH, USA), and covered with aluminum foil to keep them in dark before use. They were stored at 4 °C and were used within two weeks of preparation.

*Preparation of melted agar plates:* 3.28 g of CHROMagar™ ECC was mixed with 100 mL of reagent grade water using a magnetic stirrer bar, and the mixture was heated to 100 °C. After the mixture cooled to ~40 °C, 1 mL of bacterial suspension was mixed with the agar and dispensed into the Petri dishes. The plates were either incubated in a benchtop incubator (product no. 51030400, ThermoFisher Scientific, Waltham, MA, USA) or in our imaging platform (for monitoring the bacterial growth digitally).

We used tryptic soy agar to culture *E. coli* at 37 °C and *K. aerogenes* at 35 °C, and nutrient agar to culture *K. pneumoniae* at 37 °C. 20 g of tryptic soy agar (product no. DF0369-17-6, Fisher Scientific, Hampton, NH, USA) or 11.5 g of nutrient agar (product no. DF0001-17-0, Fisher Scientific, Hampton, NH, USA) were suspended in 500 mL of reagent grade water using a magnetic stirrer bar. The mixture was boiled on a hot plate, and then autoclaved at 121 °C for 15 min. After the mixture cooled to ~50 °C, 15 mL of the mixture was dispensed into the Petri dishes (100 mm × 15 mm) (product no. FB0875713, Fisher Scientific, Hampton, NH, USA), which were then sealed with parafilm and covered with aluminum foil to keep them in dark before use. They were stored at 4 °C until use.

*Preparation of chlorine stressed E. coli samples:* We used *E. coli* grown on tryptic soy agar plates and incubated for 48 h at 37°C. 50 mL disposable centrifuge tubes were used as a sample container and the sample size was 50 mL. 500 mL reagent grade water was filtered for sterilization using a disposable vacuum filtration unit (product no. FB12566504, Fisher Scientific, Hampton, NH, USA). A fresh chlorine suspension was prepared in a 50 mL disposable centrifuge tube to have a final concentration of 0.2 mg/mL using sodium hypochlorite (product no. 425044, Sigma Aldrich, St. Louis, MO, USA), mixed vigorously, and covered with aluminum foil.[38] 10% [w/v] sodium thiosulfate (product no. 217263, Sigma Aldrich, St. Louis, MO, USA) in reagent grade water was prepared and 1 mL of the solution was filtered using a sterile disposable syringe and a syringe filter membrane (product no. SLGV004SL, Fisher Scientific, Hampton, NH, USA) for sterilization. Water suspensions were prepared by spiking *E. coli* into filtered water samples. 50 μL of chlorine suspension (i.e. 0.2 ppm) was added to the test water sample and a timer counted the chlorine exposure time. The reaction was stopped at 10 minutes of chlorine exposure by adding 50 μL sodium thiosulfate into the test water sample and mixed vigorously immediately to stop the chlorination reaction. CHROMagar™ ECC plates were inoculated with 200 μL of chlorine stressed suspension, dried in the biosafety cabinet for at most 30 min and then placed on the setup for lens-free imaging. In addition, three TSA plates and one ECC ChromoSelect Selective Agar plate (product no. 85927, Sigma Aldrich, St. Louis, MO, USA) were inoculated with 1 mL of the control sample (not exposed to chlorine) and 0.2 ppm chlorine stressed *E. coli* water sample and dried under biosafety cabinet for about 1-2 h with gentle mixing of Petri dishes with some time intervals. After drying, the plates were sealed with parafilm and incubated at 37°C for 24 h. After the incubation, the bacterial colonies grown on the agar plates were counted and *E. coli* concentrations of control samples and the chlorine stressed *E. coli* samples were compared. If the achieved reduction in colony count was between 2.0-4.0 log, then the images of CHROMagar™ ECC plates captured using the lens-free imaging platform were used for further analysis.

*Preparation of culture plates for lens-free imaging*: A bacterial suspension in a phosphate buffer solution (PBS) (product no. 20-012-027, Fisher Scientific, Hampton, NH, USA) was prepared every day from a solid agar plate incubated for 24 h. The concentration of the suspension was measured using a spectrophotometer (model no. ND-ONE-W, Thermo Fisher), and the suspension was then diluted in the PBS to have a final concentration of 1–200 CFU per 0.1 mL. 100 µL of the diluted suspension was spread on an CHROMagar™ ECC plate using an L-shaped spreader (product no. 14-665-230, Fisher Scientific, Hampton, NH, USA). The plate was covered with its lid, inverted, and incubated at 37 °C in our optical platform (Fig. 2).

*Preparation of concentrated broth:* 180 g of tryptic soy broth (product no. R455054, Fisher Scientific, Hampton, NH, USA) was added into 1 L reagent grade water and heated to 100 °C by continuously mixing using a stirrer bar. The suspension was then cooled to 50 °C and filter sterilized using a disposable filtration unit. The broth concentrate was stored at 4 °C and used in one week after preparation.

*Preparation of samples for comparison measurements*: We evaluated the performance of our method in comparison to Colilert® 18, which is an EPA-approved enzyme based analytical method for several types of regulated water samples (e.g., drinking water, surface water, ground water) to detect *E.coli*[39] and plate counting using TSA plates and ECC ChromoSelect Selective Agar plates (Supplementary Figure S3). Two bottles of 1 L of reagent grade water were filtered using the disposable vacuum filtration units and 0.2 L of the concentrated broth was added into one of the 1 L sample bottles. The bottles covered with aluminum foil and stored in the biosafety cabinet overnight. A glass vacuum filtration unit was used for filtration of 1 L water samples. The components of the unit were covered with aluminum foil and sterilized using the autoclave. The disposable nitrocellulose filter membranes (product no. HAWG04705, EMD Millipore, Danvers, MA, USA) used in the glass filtration unit were also sterilized using the autoclave. A bacteria suspension was prepared by spiking bacteria into 50 mL reagent grade water using a disposable inoculation loop from a TSA plate containing *E. coli* colonies. The suspension was mixed gently to have uniform distribution of bacteria. Three TSA plates, 3 ECC ChromoSelect Selective Agar plates, 4 CHROMagar™ ECC plates were removed from refrigerator and kept at room temperature for 30 min.

Three bottles of 120 mL disposable vessels with sodium thiosulfate (product no. WV120SBST-200, IDEXX Laboratories Inc., Westbrook, ME, USA) were filled with 100 mL filter sterilized reagent grade water. 0.1 mL of bacterial suspension was spiked into 1 L water sample, 1.2 L water sample (1 L water + 0.2 L concentrated broth), 3 bottles of 100 mL water samples, 3 TSA plates and 3 ECC ChromoSelect Selective Agar plates, sequentially. The timer was started immediately after adding the spike into the suspensions.

First, the suspensions on TSA plates and ECC ChromoSelect Selective Agar were spread using L-shaped disposable spreaders. Then, the water sample with broth was mixed for about a minute and then stored in 35 °C for 5 h. One Colilert® 18 reagent (product no. 98-27164-00, IDEXX Laboratories Inc., Westbrook, ME, USA) was added into each 100 mL bacterial suspension and the mixture was shaken. The content of bottle was poured into a Quanti-Tray 2000 bag (product no. 98-21675-00, IDEXX Laboratories Inc., Westbrook, ME, USA) and after removing bubbles in each well, the bag was sealed using Quanti-Tray Sealer (product no. 98-09462-01, IDEXX Laboratories Inc., Westbrook, ME, USA). Three bags sealed and labelled with the experiment details were incubated at 35 °C for 18 h. Next, 30 mL filtered reagent grade water was used to moisturize the membrane in the glass filtration unit and then *E. coli* contaminated 1 L water sample was filtered at a pressure of 50 kPa. The bottle was rinsed using 150 mL of sterilized reagent grade water and the solution was filtered on the unit (Supplementary Figure S7). The funnel was rinsed using 50 mL of sterilized reagent grade water twice. After the filtration was complete, the membrane was removed and placed onto a CHROMagar™ ECC plate face down. Gentle pressure was applied on the membrane using a tweezer to remove any air bubbles between the agar and the membrane. Then, a 30 g of weight was put on the membrane to provide continuous pressure during the transfer of bacteria from the membrane to the agar plate (Supplementary Figure S8). After 5 min of incubation, the membrane was peeled off from the agar surface gently and put into another agar facing up. The agar containing the membrane was incubated at the benchtop incubator at 35 ºC and the agar containing the transferred bacteria was incubated at the lens-free imaging platform for time-lapse imaging. After 5 h of incubation, the bottle containing 1.2 L suspension was filtered using the same procedure as described before for filtration of 1 L sample. The agar plate containing the transferred bacteria was incubated at the second sample tray of the lens-free imaging setup for time-lapse imaging while the agar containing the membrane was incubated at the benchtop incubator.

b. **Design of the high-throughput time-resolved microorganism monitoring platform**

Our platform consists of five modules: (1) a holographic imaging system, (2) a mechanical translational system, (3) an incubation unit, (4) a control circuit, and (5) a controlling program. Each module is explained in detail below.

(1) We used fiber-coupled partially-coherent laser illumination (SC400-4, Fianium Ltd, Southampton, UK), with wavelength and intensity controlled through an acousto-optic tunable filter (AOTF) device (Fianium Ltd, Southampton, UK). The device is remotely controlled with a customized program written in the C++ programming language, and runs on a controlling laptop computer (product no. EON17-SLX, Origin PC). The laser light is transmitted through the sample, i.e., the agar plate that contains the bacterial colonies, and forms an inline hologram on a CMOS image sensor (product no. acA3800-14um, Basler AG, Ahrensburg, Germany) with a pixel size of 1.67 μm and an active area of 6.4 mm × 4.6 mm. The CMOS image sensor is connected to the same controlling laptop computer through a universal serial bus (USB) 3.0 interface and is software-triggered within the same C++ program. The exposure time at each scanning position is pre-calibrated according to the intensity distribution of the illumination light, and ranges from 4 ms to 167 ms. The images are saved as 8-bit bitmap files for further processing.

(2) The mechanical stage is customized with a pair of linear translation rails (Accumini 2AD10AAAHL, Thomson, Radford, VA, USA), a pair of linear bearing rods (8 mm diameter, generic), and linear bearings (LM8UU, generic), and it is aided by parts printed by a 3D printer for the joints and housing (Objet30 Pro, Stratasys, Minnesota, USA). The two-dimensional horizontal movement is powered by two stepper motors (product no. 1124090, Kysan Electronics, San Jose, CA, USA)—one for each direction, and these motors are individually controlled using stepper motor controller chips (DRV8834, Pololu Las Vegas, NV, US). To minimize the backslash effect, the whole Petri dish is scanned following the raster scan pattern.

(3) The incubation unit is built with the top heating plate of a microscope incubator (INUBTFP-WSKM-F1, Tokai Hit, Shizuoka, Japan), and it is housed by a 3D frame printed by a 3D printer. The Petri dish containing the sample is placed on the heating plate with the surface having bacteria facing downwards. The temperature is controlled by a paired controller that maintains a temperature of 47 °C on the heating plate, resulting in a temperature of 38 °C inside the Petri dish.

(4) The control circuit consists of three components: a micro-controller (Arduino Micro, Arduino LLC) communicating with the computer through a USB 2.0 interface, two stepper motor driver chips (DRV8834, Pololu Las Vegas, NV, US) externally powered by a 4.2 V constant voltage power supply (GPS-3303, GW Instek, Montclair, CA, US), and a metal–oxide–semiconductor field-effect transistor (MOSFET)-based digital switch (SUP75P03-07, Vishay Siliconix, Shelton, CT, United States) for controlling the CMOS sensor connection.

(5) The controlling program includes a graphical user interface (GUI) and was developed using the C++ programming language. External libraries including Qt (v5.9.3), AOTF (Gooch & Housego), and Pylon (v5.0.11) were integrated.

c. **Data acquisition**

We prepared inoculated agar plates of pure bacterial colonies (see the Sample Preparation subsection under the Methods for details), and captured images of an entire agar plate at 30-minute intervals. The illumination light was set to a wavelength of 532 nm and an intensity of ~400 μW. To maximize the image acquisition speed, the captured images were first saved into a computer memory buffer and then written to hard disk by another independent thread. At the end of each experiment (i.e., after 24 h of incubation), the sample plate was imaged using a benchtop scanning microscope (Olympus IX83) in reflection mode, and the resulting images were automatically stitched to a full-FOV image, used for comparison. Subsequently, the plate was disposed of as solid biohazardous waste. We populated data (i.e., time-lapse lens-free images) corresponding to ~6,969 *E. coli*, ~2,613 *K. aerogenes*, and ~6,727 *K. pneumoniae* individual bacterial colonies to train and validate our models. Another 965 colonies of 3 different species from 15 independent agar-plates were used to blindly test our machine learning models.

d. **Image processing and analysis**

The acquired lens-free images are processed using custom-developed image processing and deep learning algorithms. There are five major image processing steps for the early detection and automated classification and counting of colonies. These steps are described in detail below.

i. *Image stitching to obtain the image of the entire plate area:* Following the acquisition of holographic images using the multi-threading approach, all the images within a tile-scan of the whole Petri dish per wavelength are merged into a single full-FOV image. During a tile scan, the images are acquired with ~30% overlap on each side of the image, to calculate the relative image shifts against each other. As for each image, the relative shifts against all four of the neighboring images are calculated using a phase correlation[40] method, followed by an optimization step that minimizes an object function, as defined by:

$$\arg\min_{T_{VF}} \sum_{A \in V \setminus \{F\}} \left( \sum_{B \in V \setminus \{F\}} \left\| \vec{t}_{AF} - \vec{t}_{BF} - \vec{p}_{AB} \right\|^2 \right), \tag{1}$$

where $V$ is the set of all tile images, $F \in V$ is a fixed image, e.g., the image captured at the center of the sample Petri dish, $\vec{t}_{AB}$ stands for the relative position of image $A$ with respect to image $B$, and $\vec{p}_{AB}$ is the local shift between images $A$ and $B$, calculated by the phase correlation method using the overlapping regions of the two neighboring images, which can be formulated as:

$$\vec{p}_{AB} = (\Delta x, \Delta y) = \arg\max_{(x,y)} \mathcal{F}^{-1}\left\{ \frac{\mathcal{F}\{A\} \cdot \mathcal{F}\{B\}^*}{|\mathcal{F}\{A\} \cdot \mathcal{F}\{B\}^*|} \right\} \qquad (2)$$

where $\mathcal{F}$ is the Fourier transform operator and $\mathcal{F}^{-1}$ is the inverse Fourier transform operator. The optimal configuration $T_{VF} = \{\vec{t}_{AF} : A, F \in V\}$ represents the relative positions of all the images with respect to the fixed image $F$ and it is used as the global position of each tile image for full-FOV image stitching. To eliminate tiles with a low signal-to-noise ratio (SNR) that lead to incorrect local shift estimation values, a correlation threshold of 0.3 is applied during the optimization, meaning that if the cross-correlation coefficient of the overlapped parts of two images is below 0.3, the shift calculation is discarded. Once the positions of all of the tiles are obtained, they are merged into a full FOV image of the whole Petri dish using linear blending. We define a full-FOV image of the whole Petri dish as a 'frame'. All the frames are normalized so that the mean value is 50, and they are saved as unsigned 8-bit integer (0-255) arrays.

ii. *Colony candidate selection by differential analysis*: When a new frame is acquired at time $t$, it is cross-registered to the previous frame at time $t-1$, and then is digitally back-propagated to the sample plane[41,42] to obtain the complex light field

$$\tilde{B}_t = \mathrm{P}(F_t, \mathbf{z}), \qquad (3)$$

where $F_t$ is the frame at time $t$, $\mathbf{z}$ is a surface normal vector of the sample plane obtained by digital auto-focusing[43] at 50 randomly-spaced positions, and P denotes the angular spectrum-based back-propagation operation.[41,42] To accommodate the large FOV of a stitched frame (36000 × 36000 pixels), digital back-propagation is performed with 2048 × 2048-pixel blocks, which are then merged together.

We take four consecutive frames, i.e., from $t-3$ to $t$, and calculate a differential image defined by:

$$D_t = \mathrm{HP}\left[ \mathrm{LP}\left( \frac{1}{3} \sum_{\tau=t-2}^{t} |\tilde{B}_\tau - \tilde{B}_{\tau-1}| \right) \right], \qquad (4)$$

where $D_t$ is the differential image at time $t$, $\tilde{B}_t$ represents the complex light field obtained by back-propagating frame $t$, and LP and HP represent low-pass and high-pass image filtering, respectively. The HP filter removes the differential signal from a slowly-varying background (unwanted term), and the LP filter removes the high-frequency noise-introduced spatial patterns. The LP and HP filter kernels are empirically set to 5 and 100, respectively.

Following the differential image calculation, we select regions in the differential image with >50 connective pixels that are above an intensity threshold, which is empirically set to 12. These regions are marked as colony candidates, as they give a differential signal over a period of time (covering four consecutive frames). However, some of the differential signal comes from non-bacterial objects, such as a water bubble or surface movement of the agar itself. Therefore, we also use two DNNs to select the true candidates and classify their species.

iii. *DNN-enabled detection of growing bacterial colonies*: Following the colony candidate selection process outlined earlier, we crop out candidate regions of $160 \times 160$ pixels (~$267 \times 267$ μm$^2$) across the four back-propagated consecutive frames and separate the complex field into amplitude and phase channels. Therefore, each candidate region is represented by a $2 \times 4 \times 160 \times 160$ array. This four-dimensional (phase/amplitude-time-x-y) data format differs from the traditional three-dimensional data used in image classification tasks and requires a custom-designed DNN architecture that accounts for the additional dimension of time. We designed our DNN by following the block diagram of DenseNet[25], and replaced the 2D convolutional layers with P3D convolutional layers[44], as shown in Supplementary Fig. S9. Our network was implemented in Python (v3.7.2) with the PyTorch Library (v1.0.1). The network was randomly initialized and optimized using an adaptive moment estimation (Adam) optimizer[45] with a starting learning rate of $1 \times 10^{-4}$ and a batch size of 64. To stabilize the accuracy of the network model, we also set a learning rate-scheduler that decayed the learning rate by half every 20 epochs. Approximately 16,000 growing colonies and 43,000 non-colony objects captured from 71 agar plates were used in the training and validation phases. The best network model was selected based on the best validation accuracy. Data augmentation was also applied by random 90°-rotations and flipping operations in the spatial dimensions. The whole training process took ~5 h using a desktop computer with dual GPUs (GTX1080Ti, Nvidia). The decision threshold value after the softmax layer was set to 0.5 during training, i.e., positive for softmax value >0.5 and negative for softmax value <0.5, which implies equal penalty to false positive and false negative events. We adjusted the threshold value to 0.99, empirically based on the training dataset before blind testing, to favor less false positive events.

iv. *DNN-enabled classification of bacterial colony species*: Once the true bacterial colonies are selected, they grow for another 2 h to collect 8 consecutive frames, i.e., 4 h, and then are sent to the second DNN as a $2 \times 8 \times 288 \times 288$ array for classification of colony species. To perform the classification task, this time, the training data only contain the true colonies and their corresponding species (ground truth). The network follows a similar structure and training process as the detection model, as illustrated in Supplementary Fig. S9. The network was randomly initialized and optimized using the Adam optimizer[45], with a starting learning rate of $1 \times 10^{-4}$ and a batch size of 64. The learning rate decayed by 0.9 times every 10 epochs. To avoid overfitting to a specific plate, we discarded colony images extracted from extremely dense samples (>1000 CFU per plate). As a result, approximately 9,400 growing colonies were used in the training and validation of the classification model. The whole training process took ~15 h using a desktop computer with dual GPUs (GTX1080Ti, Nvidia).

e. *Colony counting:* The respective ground truth information on the growing colonies in each experiment was created after the sample is incubated for >24 h. At the boundary of the plate, the agar always forms a curved surface owing to surface tension, thereby distorting the images of the colonies. Therefore, we limit the effective imaging area to a 50 mm-diameter circle in the center of the agar-plate. In the cases, where multiple colonies are closely spaced and eventually merge into one large colony (e.g., toward the end of 24 h incubation period), we then use the lens-free time-lapsed images to verify the true colony number when detected by our method, so as to avoid over-counting.

f. **Calculation of imaging throughput**

In Supplementary Table S2, we compared the imaging throughput of our system and a conventional lens-based scanning microscope in terms of the space-bandwidth product (SBP)[46] using the following formula:

$$N_I = \alpha \cdot \text{FOV} \cdot r^2 / \delta^2 \qquad (5)$$

where $N_I$ is the effective pixel-count of a frame, $\delta$ is the half-pitch resolution, $r$ is the digital sampling factor along the $x$ and $y$ directions, and $\alpha = 2$ represents the independent spatial information contained in the phase and amplitude images of the holographic reconstruction, while $\alpha = 1$ represents the amplitude-only information contained in an image captured using the standard lens-based bright-field scanning microscope. In the lens-based microscope, we used a color camera with a pixel size of 7.4 µm. Therefore, for a 4× objective lens the image resolution is limited to ~3.7 µm, owing to the Nyquist sampling limit. Without loss of generality, we set $r = 2$.[47]

*Acknowledgments*: The authors acknowledge the funding of ARO (Contract # W911NF-17-1-0161), Koc Group and HHMI. The authors would also like to acknowledge IDEXX Laboratories Inc. for loaning the Quanti-Tray Sealer and Drs. Janine R. Hutchison and Richard M. Ozanich from Pacific Northwest National Laboratory for sharing their assistance on the chlorination of bacteria samples.

**References**

1. Sandgren, A. *et al.* Tuberculosis drug resistance mutation database. *PLoS medicine* 6, e1000002 (2009).
2. Arain, T. M., Resconi, A. E., Hickey, M. J. & Stover, C. K. Bioluminescence screening in vitro (Bio-Siv) assays for high-volume antimycobacterial drug discovery. *Antimicrobial agents and chemotherapy* 40, 1536–1541 (1996).
3. Jacobs, W. R. *et al.* Rapid assessment of drug susceptibilities of Mycobacterium tuberculosis by means of luciferase reporter phages. *Science* 260, 819–822 (1993).
4. Goodacre, R. *et al.* Rapid identification of urinary tract infection bacteria using hyperspectral whole-organism fingerprinting and artificial neural networks. *Microbiology* 144, 1157–1170 (1998).
5. Lagier, J.-C. *et al.* Culturing the human microbiota and culturomics. *Nature Reviews Microbiology* 1 (2018).
6. Fierer, N. *et al.* Forensic identification using skin bacterial communities. *Proceedings of the National Academy of Sciences* 107, 6477–6481 (2010).
7. Koydemir, H. C. *et al.* Rapid imaging, detection and quantification of Giardia lamblia cysts using mobile-phone based fluorescent microscopy and machine learning. *Lab on a chip* 15, 1284–1293 (2015).
8. Oliver, S. P., Jayarao, B. M. & Almeida, R. A. Foodborne pathogens in milk and the dairy farm environment: food safety and public health implications. *Foodbourne Pathogens & Disease* 2, 115–129 (2005).
9. World Water Day. https://www.cdc.gov/healthywater/observances/wwd.html?CDC_AA_refVal=https%3A%2F%2Fwww.cdc.gov%2Ffeatures%2Fworldwaterday%2Findex.html.

10. DeFlorio-Barker, S., Wing, C., Jones, R. M. & Dorevitch, S. Estimate of incidence and cost of recreational waterborne illness on United States surface waters. *Environmental Health* 17, 3 (2018).
11. *Method 1604: total coliforms and Escherichia coli in water by membrane filtration using a simultaneous detection technique (MI Medium)*. (United States, Environmental Protection Agency, Office of Water, 2002).
12. Current Waterborne Disease Burden Data & Gaps | Healthy Water | CDC. https://www.cdc.gov/healthywater/burden/current-data.html (2018).
13. US EPA. Analytical Methods Approved for Compliance Monitoring under the Long Term 2 Enhanced Surface Water Treatment Rule. (2017).
14. Deshmukh, R. A., Joshi, K., Bhand, S. & Roy, U. Recent developments in detection and enumeration of waterborne bacteria: a retrospective minireview. *MicrobiologyOpen* 5, 901–922 (2016).
15. Amann, R. & Fuchs, B. M. Single-cell identification in microbial communities by improved fluorescence *in situ* hybridization techniques. *Nature Reviews Microbiology* 6, 339–348 (2008).
16. Kang, D.-K. *et al.* Rapid detection of single bacteria in unprocessed blood using Integrated Comprehensive Droplet Digital Detection. *Nature communications* 5, 5427 (2014).
17. *Title 40: Protection of Environment. Electronic Code of Federal Regulations* vol. 136.3.
18. Van Poucke, S. O. & Nelis, H. J. A 210-min solid phase cytometry test for the enumeration of Escherichia coli in drinking water. *Journal of applied microbiology* 89, 390–396 (2000).
19. Kim, M. *et al.* Optofluidic ultrahigh-throughput detection of fluorescent drops. *Lab on a Chip* 15, 1417–1423 (2015).
20. Tryland, I., Braathen, H., Wennberg, A., Eregno, F. & Beschorner, A.-L. Monitoring of β-D-Galactosidase activity as a surrogate parameter for rapid detection of sewage contamination in urban recreational water. *Water* 8, 65 (2016).
21. Van Poucke, S. O. & Nelis, H. J. Limitations of highly sensitive enzymatic presence-absence tests for detection of waterborne coliforms and Escherichia coli. *Appl. Environ. Microbiol.* 63, 771–774 (1997).
22. London, R. *et al.* An Automated System for Rapid Non-Destructive Enumeration of Growing Microbes. *PLOS ONE* 5, e8609 (2010).
23. *EPA Microbiological Alternate Test Procedure (ATP) Protocol for Drinking Water, Ambient Water, Wastewater, and Sewage Sludge Monitoring Methods*. (United States, Environmental Protection Agency, Office of Water, 2010).
24. CHROMagar$^{TM}$ ECC Product Leaflet. http://www.chromagar.com/fichiers/1559127431LF_EXT_003_EF_V8.0.pdf?PHPSESSID=37ddd615300a5b1a756549fa91cdb437
25. Huang, G., Liu, Z., van der Maaten, L. & Weinberger, K. Q. Densely Connected Convolutional Networks. *arXiv:1608.06993 [cs]* (2016).
26. Shapiro, J. A. The significances of bacterial colony patterns. *BioEssays* 17, 597–607 (1995).
27. Su, P.-T. *et al.* Bacterial Colony from Two-Dimensional Division to Three-Dimensional Development. *PLOS ONE* 7, e48098 (2012).
28. Farrell Fred D., Gralka Matti, Hallatschek Oskar & Waclaw Bartlomiej. Mechanical interactions in bacterial colonies and the surfing probability of beneficial mutations. *Journal of The Royal Society Interface* 14, 20170073 (2017).
29. Sheats Julian, Sclavi Bianca, Cosentino Lagomarsino Marco, Cicuta Pietro & Dorfman Kevin D. Role of growth rate on the orientational alignment of Escherichia coli in a slit. *Royal Society Open Science* 4, 170463.
30. LeChevallier, M. W. & McFeters, G. A. Enumerating Injured Coliforms in Drinking Water. *Journal (American Water Works Association)* 77, 81–87 (1985).
31. CDC-Salmonella-Factsheet. https://www.cdc.gov/salmonella/pdf/CDC-Salmonella-Factsheet.pdf.
32. Liu, H., Whitehouse, C. A. & Li, B. Presence and Persistence of Salmonella in Water: The Impact on Microbial Quality of Water and Food Safety. *Front Public Health* 6, (2018).

33. Alternate Test Procedures in Clean Water Act Analytical Methods. https://www.epa.gov/cwa-methods/alternate-test-procedures.
34. Sanders, E. R. Aseptic Laboratory Techniques: Plating Methods. *J Vis Exp* (2012) doi:10.3791/3064.
35. Zhang, Y. *et al.* Motility-based label-free detection of parasites in bodily fluids using holographic speckle analysis and deep learning. *Light: Science & Applications* 7, 108 (2018).
36. Isikman, S. O. *et al.* Lens-free optical tomographic microscope with a large imaging volume on a chip. *PNAS* 108, 7296–7301 (2011).
37. Cobo, M. P. *et al.* Visualizing bacterial colony morphologies using time-lapse imaging chamber MOCHA. *Journal of bacteriology* 200, e00413-17 (2018).
38. Hutchison, J. R. *et al.* Consistent production of chlorine-stressed bacteria from non-chlorinated secondary sewage effluents for use in the U.S. Environmental Protection Agency Alternate Test Procedure protocol. *Journal of Microbiological Methods* 163, 105651 (2019).
39. Colilert 18 - IDEXX US. https://www.idexx.com/en/water/water-products-services/colilert-18/.
40. Preibisch, S., Saalfeld, S. & Tomancak, P. Globally optimal stitching of tiled 3D microscopic image acquisitions. *Bioinformatics* 25, 1463–1465 (2009).
41. Goodman, J. W. *Introduction to Fourier Optics*. (Roberts and Company Publishers, 2005).
42. Greenbaum, A. *et al.* Wide-field computational imaging of pathology slides using lens-free on-chip microscopy. *Science Translational Medicine* 6, 267ra175-267ra175 (2014).
43. Zhang, Y., Wang, H., Wu, Y., Tamamitsu, M. & Ozcan, A. Edge sparsity criterion for robust holographic autofocusing. *Opt. Lett., OL* 42, 3824–3827 (2017).
44. Qiu, Z., Yao, T. & Mei, T. Learning Spatio-Temporal Representation with Pseudo-3D Residual Networks. *arXiv:1711.10305 [cs]* (2017).
45. Kingma, D. P. & Ba, J. Adam: A Method for Stochastic Optimization. *arXiv:1412.6980 [cs]* (2014).
46. Wang, H. *et al.* Computational out-of-focus imaging increases the space–bandwidth product in lens-based coherent microscopy. *Optica, OPTICA* 3, 1422–1429 (2016).
47. Greenbaum, A. *et al.* Increased space-bandwidth product in pixel super-resolved lensfree on-chip microscopy. *Scientific Reports* 3, (2013).

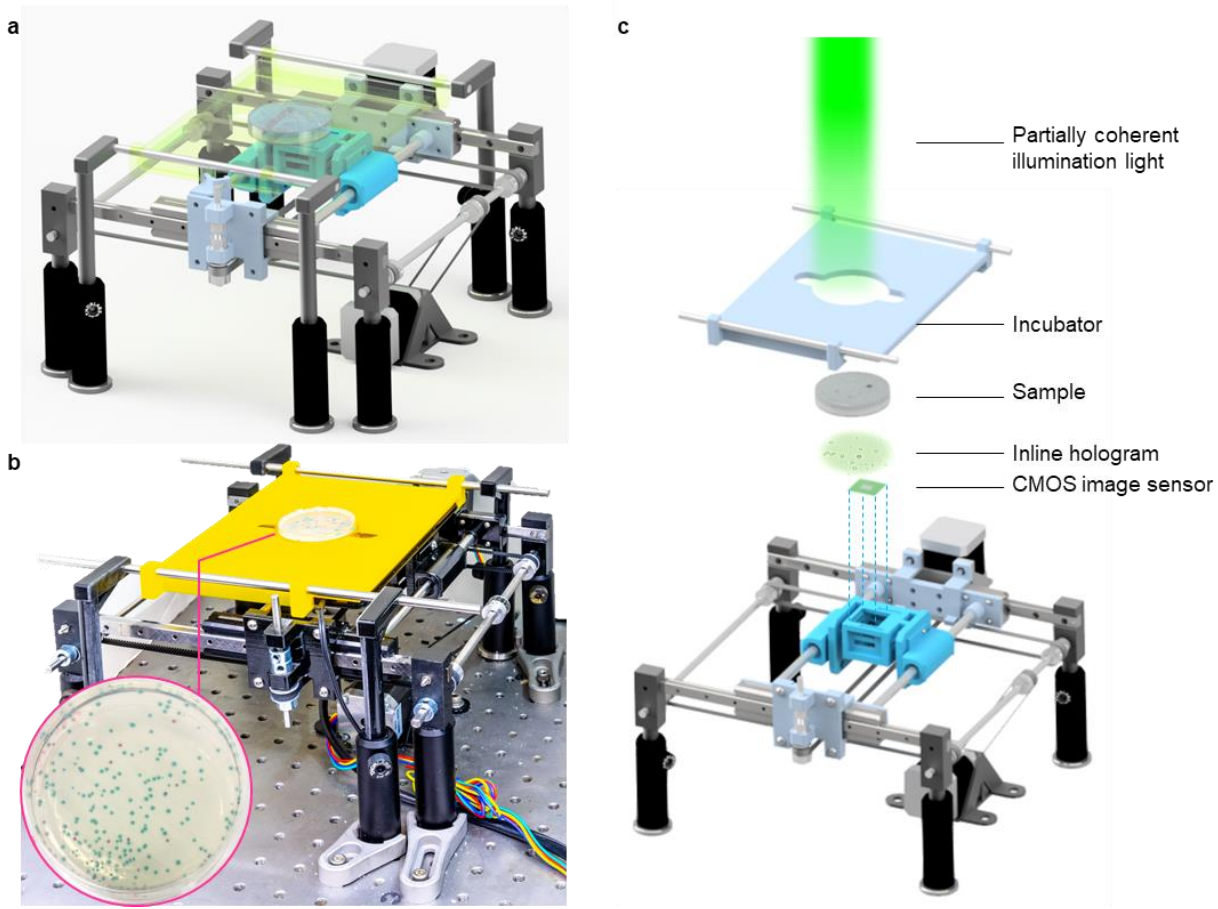

**Figure 1.** High-throughput bacterial colony growth detection and classification system. (a) Schematic of the device. (b) Photograph of the lens-free imaging system. (c) Detailed illustration of various components of the system.

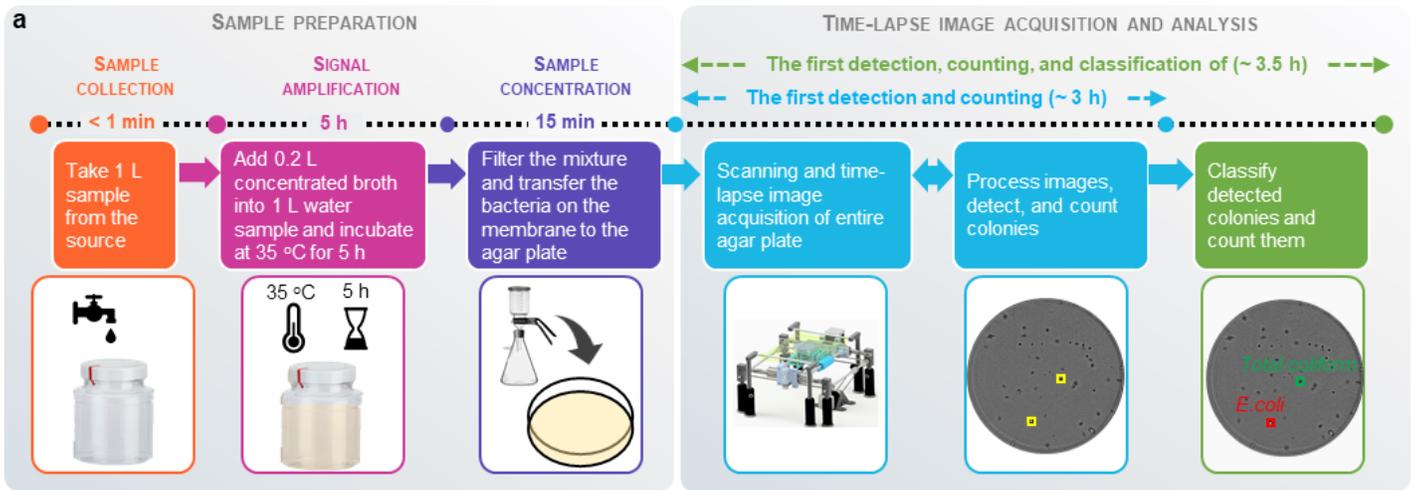
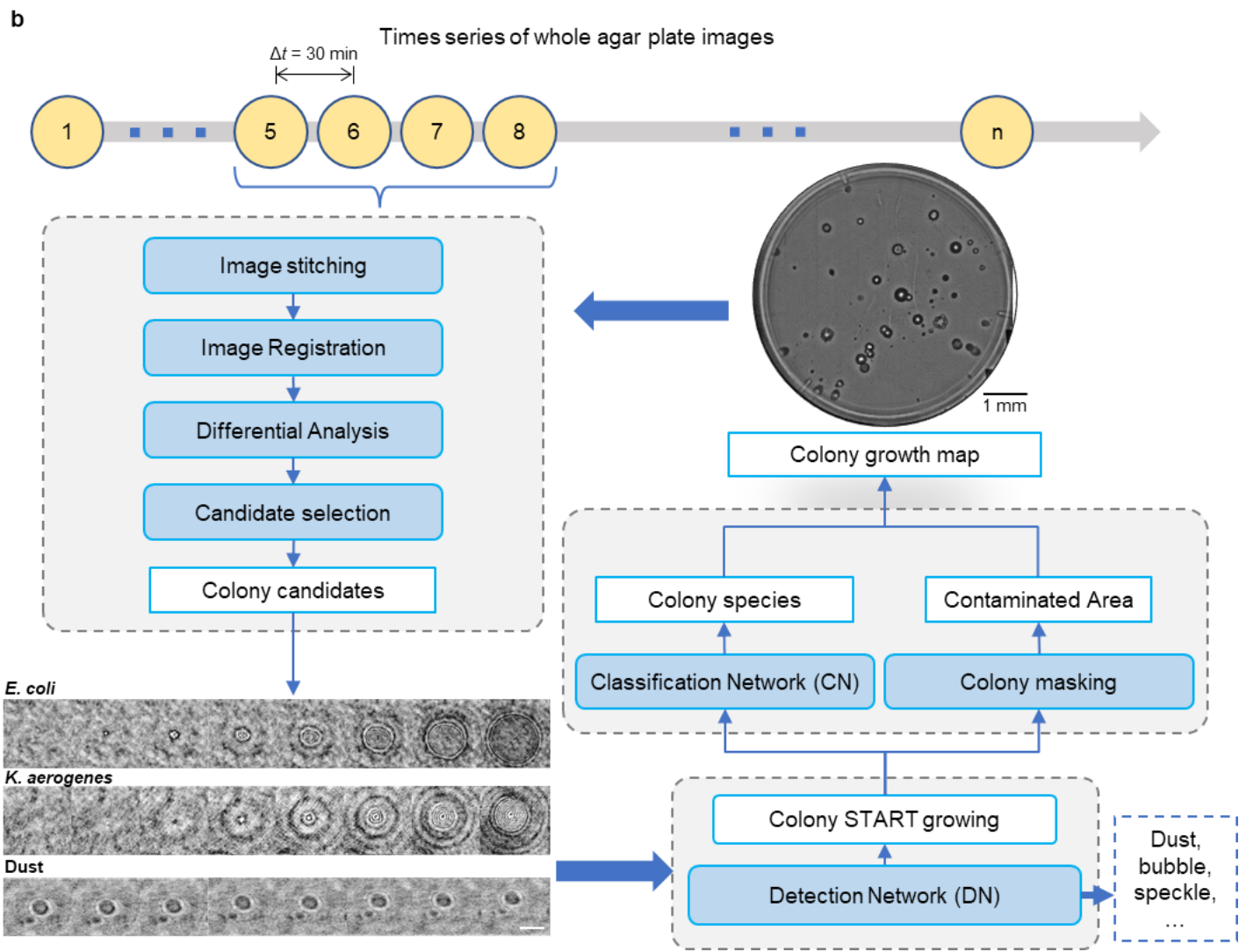

**Figure 2.** Schematics demonstrating the workflow of the microorganism monitoring system. (a) Bacterial sample preparation workflow. (b) Steps of the image and data processing algorithms for automated detection of growing colonies and classification of their species. The scale bars for the holographic images of the growing colonies (*E. coli* and *K. aerogenes*) and a static particle (dust) are 100 μm.

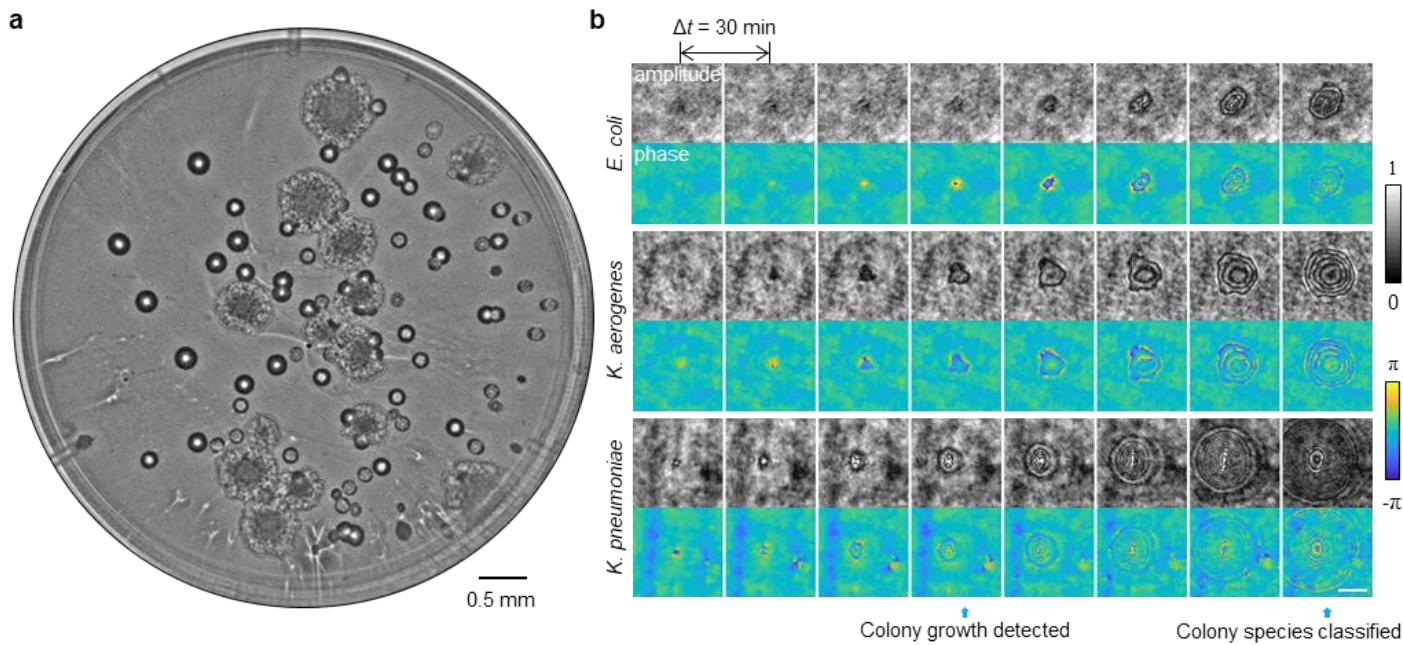

**Figure 3.** Images captured using the microorganism monitoring system. (a) Whole agar-plate image of mixed *E. coli* and *K. aerogenes* colonies, after 23.5 h of incubation. (b) Example images (i.e., amplitude and phase) of the individual growing colonies detected by a trained deep neural network. The time points of detection and classification of growing colonies are annotated with blue arrows. The scale bar is 100 μm.

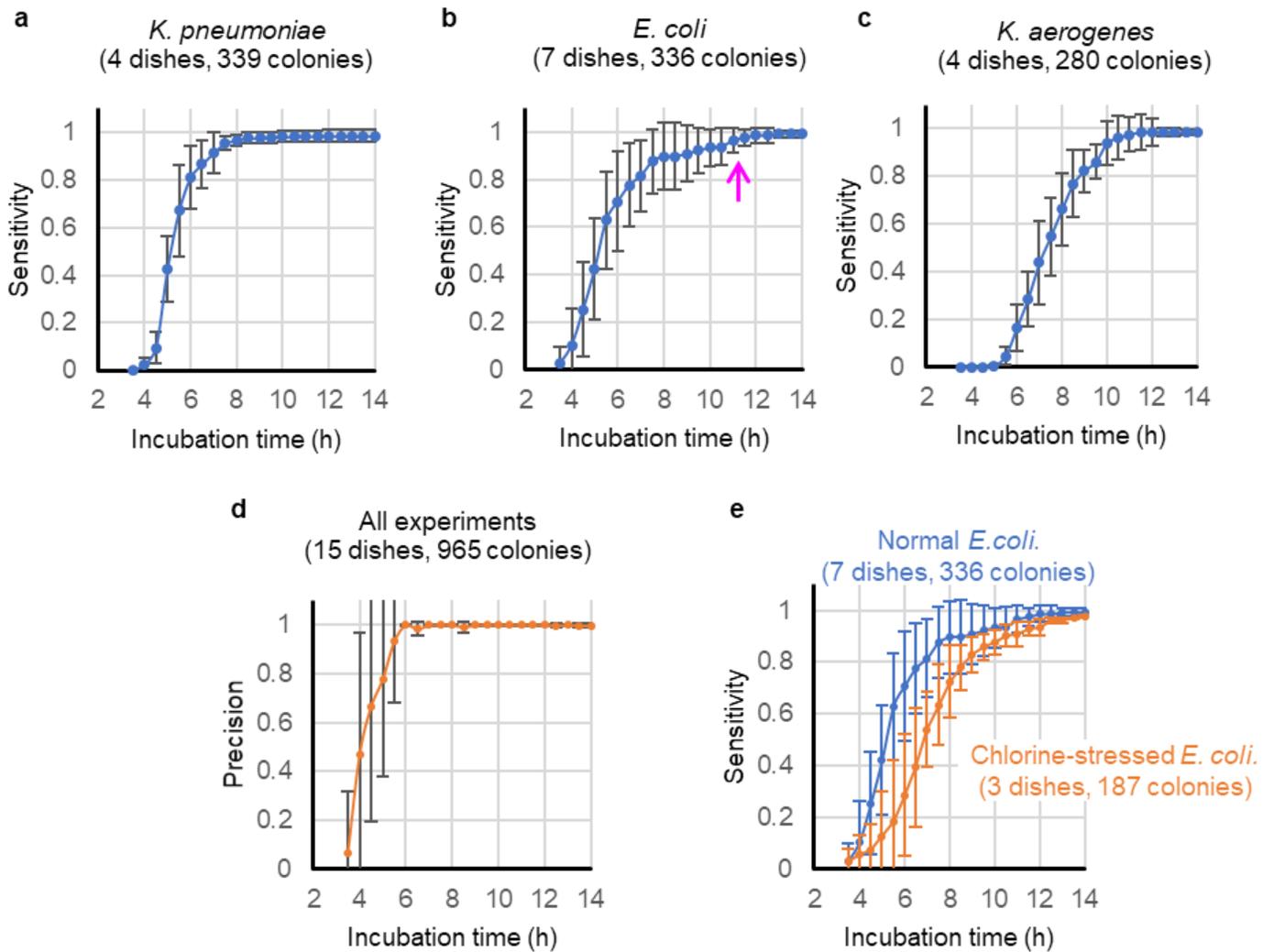

**Figure 4.** Sensitivity of growing colony detection using our trained neural network for (a) *K. pneumoniae,* (b) *E. coli*, and *K. aerogenes*. (d) Precision of growing colony detection using our trained neural network for all three species. Pink arrow indicates the time for late "wake-up" behavior for some of the *E. coli* colonies. (e) Characterizing the growth speed of chlorine-stressed *E. coli* using our system. There is a ~2 h delay of colony formation for chlorine-stressed *E. coli* (the orange curve) compared to the unstressed *E. coli* strain (the blue curve).

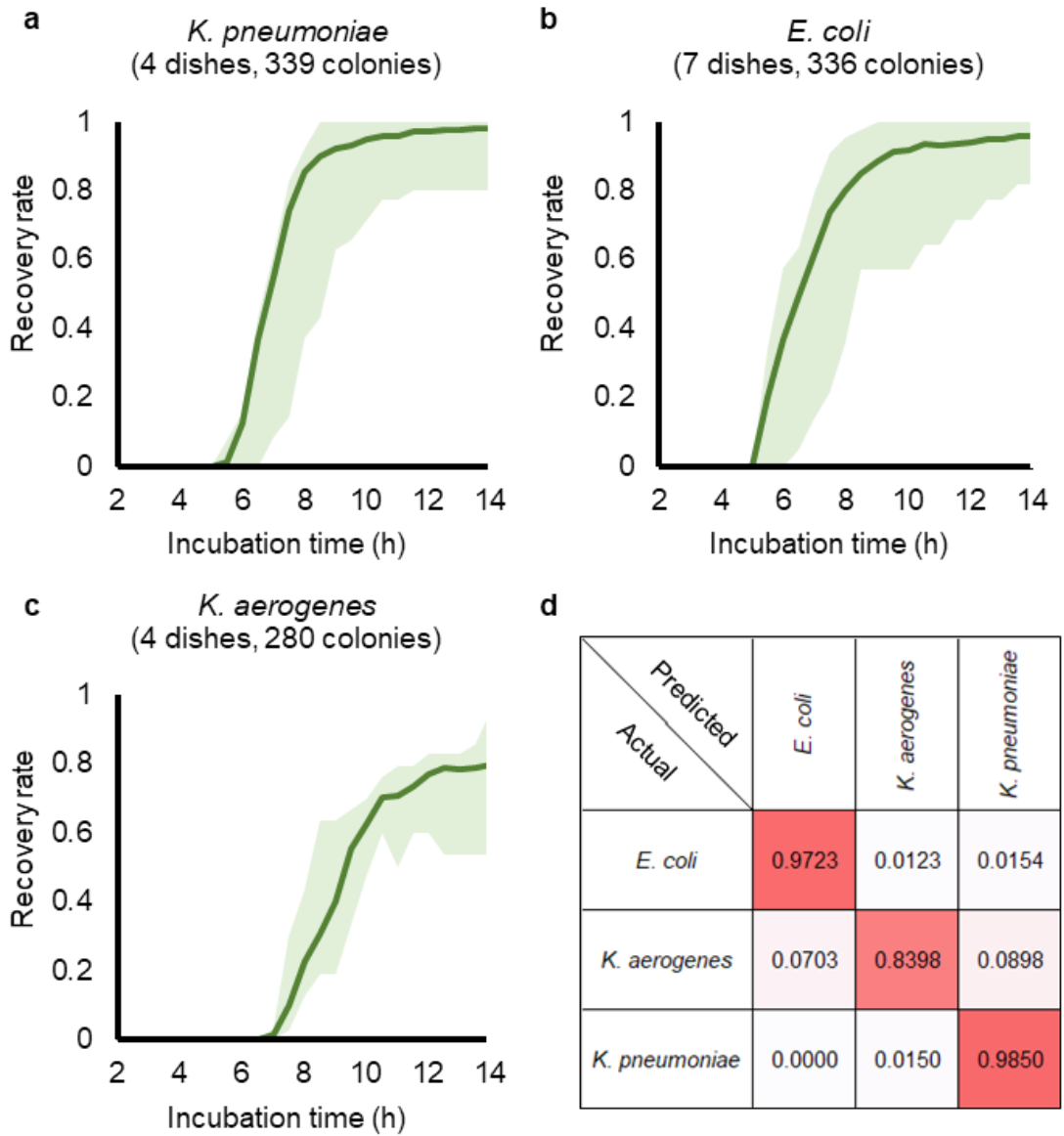

**Figure 5.** Classification performance of our trained neural network for (a) *K. pneumoniae*, (b) *E. coli*, and (c) *K. aerogenes* colonies. The green shaded area in each curve represents the highest and lowest recovery rates found in all the corresponding experiments at each time point. (d) The blind testing confusion matrix of classifying all the colonies that were sent to our trained neural network after 12 h our incubation. A diagonal entry of 1.0 means 100% classification accuracy for that species. The number of colonies that were tested by the classification network in (d): 325 (*E. coli*), 334 (*K. pneumoniae*), and 256 (*K. aerogenes*).

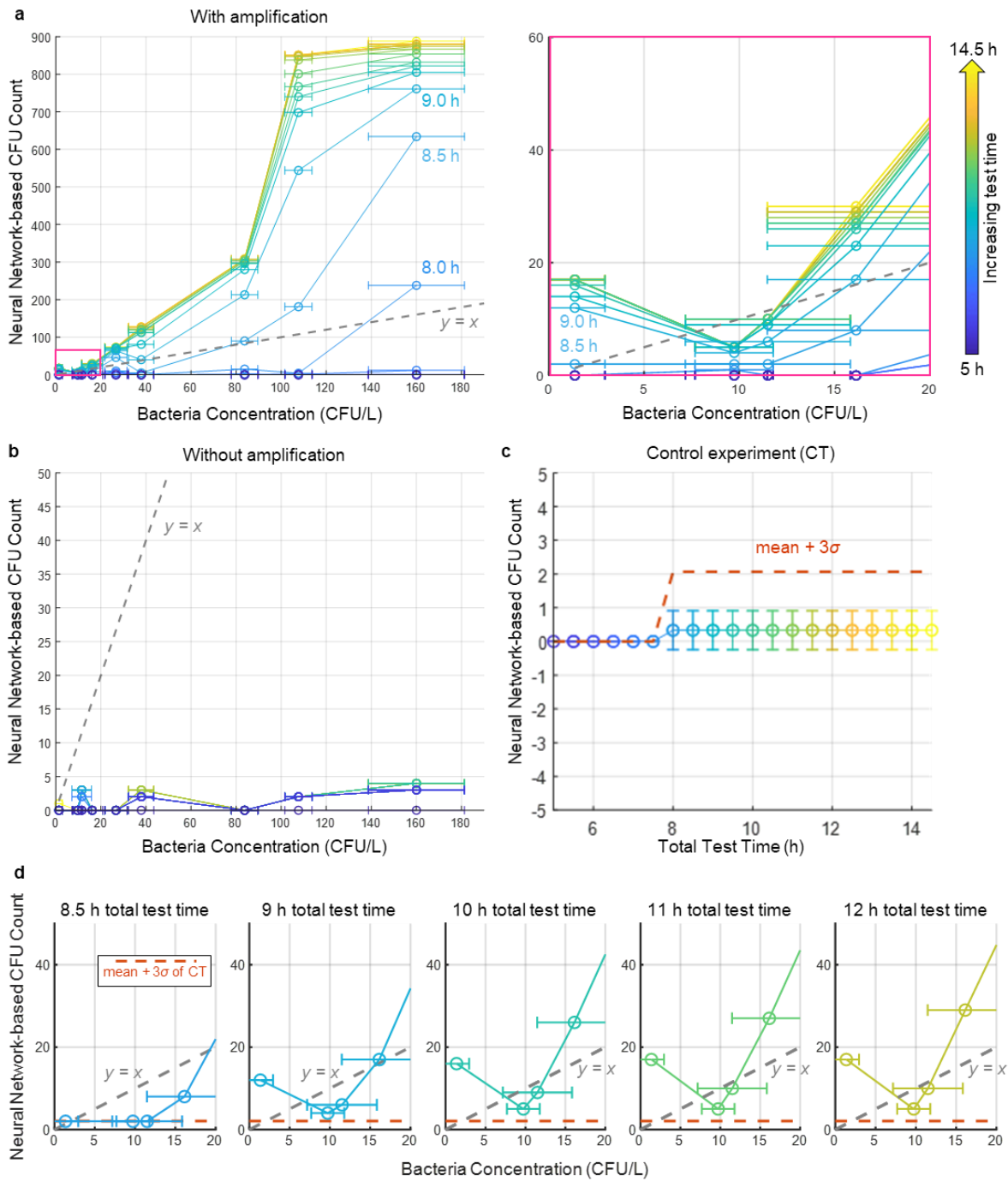

**Figure 6**. Quantification of the LOD of our system. (a) The CFU count from our system is plotted against the CFU/L counts of the spiked samples, calculated independently using the Colilert®18 method after 18 hours of incubation. CFU counts acquired with our platform at different time points are colored from blue to yellow, which corresponds to 5 to 14.5 hours of total test time, including the signal amplification step that involves liquid culture media (5 h). (b) Without signal amplification, the LOD is decreased due the low transfer rate from filter membrane to the agar surface (see Supplementary Figures S3 and S4). (c) As a control experiment, we prepared and imaged 3 agar plates which show <1 CFU count from our setup throughout the test period from 5 h to 14.5 h. (d)The LOD of our system is ~11 CFU/L at 8.5 h and ~1 CFU/L at ≤9 h.